\documentclass[showpacs,preprintnumbers,amsmath,amssymb]{revtex4}

\usepackage{graphicx}

\begin{document}

\title{Hybrid molecular-continuum fluid dynamics}

\author{Rafael Delgado-Buscalioni}
\email{R.Delgado-Buscalioni@ucl.ac.uk}
\affiliation{Centre for Computational Science, Department of Chemistry  \\
University College London,
\\  20 Gordon Street, London WC1 OAJ, UK}
\author{Peter V. Coveney}
\email{p.v.coveney@ucl.ac.uk}
\affiliation{Centre for Computational Science, Department of Chemistry  \\
University College London,
\\  20 Gordon Street, London WC1 OAJ, UK}

\label{firstpage}

\begin{abstract}
We describe recent developments in the hybrid atomistic/continuum modelling of
dense fluids. We discuss the general implementation of 
mass, momentum and energy transfers between a region described by molecular
dynamics and the neighbouring domain described by the Navier-Stokes
equations for unsteady flow.
\end{abstract}

\maketitle
\section{Introduction}

The flow of complex fluids near interfaces is governed by a subtle interplay between the
fast microscopic dynamics within a small localised region of the
system close to the interface and the slow dynamics in the bulk fluid region. 
This scenario is encountered in a wide variety of applications
ranging from nanotechnology (nanofluidics) and other 
industrial processes (such as wetting, droplet formation, critical fluids near heated surfaces or crystal growth
from a fluid phase) to biological systems (for example, membranes or biomolecules near interfaces).
The dynamics of these systems depends on the intimate connection of 
many different spatio-temporal scales: from the nanoscale to the
microscale and beyond. 
Realistic simulations of such systems via standard classical molecular dynamics
(MD) are prohibitive, while continuum fluid dynamics (CFD) cannot
describe the important details within the interfacial region. 
In view of this fact, the field of computer 
simulation is now faced with the need for new techniques, which bridge
a wider range of time and length scales with the minimum loss of information.
A hybrid particle-continuum approach provides a resolution to this dilemma. 
A hybrid algorithm retains all the atomistic detail within the relevant
localized domain and couples this region to 
the continuum hydrodynamic description of the remainder of the system.
Indeed hybrid algorithms for liquids can be expected to provide a powerful tool for the fast
growing field of nanofluidics in micro electro-mechanical systems
(MEMS) and our ongoing contributions have been recognized by the nanoscience
community (R. Delgado-Buscalioni and Coveney 2003a) as offering a promising simulation
technique with nanotechnological applications.

Hybrid algorithms for solids (Abraham {\em et al.} 1998) and
gases (Garcia {\em et al} 1999) were the first to be fully developed in the
literature.  As expected in most theoretical descriptions of matter, the
hybrid description of the liquid state is the most challenging one.
The general procedure is to connect the particle domain (P) and the
continuum domain (C) 
within an overlapping region comprised of two buffers: C$\rightarrow$P and
P$\rightarrow$C (see   figure  1).   Within  the
P$\rightarrow$C   buffer  the particle dynamics are coarse-grained
to  extract the boundary conditions for the  C-region.  
The  most complicated part
of any hybrid scheme is the C$\rightarrow$P coupling
where the microscopic dynamics need to be
reconstructed to adhere to the prescriptions given by the  continuum
variables. Moreover, in doing so the  unphysical  artifacts thereby
introduced should  be minimized (following Occam's razor).

In this paper we provide an overview of the state-of-the-art of the hybrid
modelling of liquids. In \S \ref{s-o} we start by presenting an
overview of the hybrid scheme and some preliminary
topics such as the inherent constraints on the continuum time
step and the spatial-grid size. Section \ref{t-c} discusses several
implementations of the temporal-coupling. The C$\rightarrow$P
coupling scheme is then explained in \S \ref{s-cp} for the general case of
mass, momentum and energy.  We illustrate this important part of the
scheme by reproducing the three hydrodynamic modes (shear, sound and
heat) governing the relaxing flows in an infinite medium. Section
\ref{s-pc} is devoted to the P$\rightarrow$C coupling, based  on a 
finite volume method solving the flow within
the C domain.  Some comments on the effect of noise on the accuracy of
the scheme are made. The full method is used in \S \ref{osc}
to solve the problem of shear flow driven by oscillatory wall motion in a
nano-slot. Finally, conclusions and future directions for this research
are described in \S \ref{cn}.

\section{Overview
\label{s-o}}
The domain decomposition deployed in our hybrid scheme is depicted in
figure 1.  Within domain P the fluid is described at the atomistic
level via Newtonian dynamics.  The position of the $N(t)$ atoms at
time $t$ inside P is updated each $\Delta t_P$ time interval using a
standard MD scheme.  The present calculations were done with a
Lennard-Jones (LJ) fluid.  Throughout the ongoing discussion all
quantities are given in reduced Lennard Jones units: length $\sigma$,
mass $m$, energy $\epsilon$, time $(m\sigma^2/\epsilon)^{1/2}$ and
temperature $\epsilon/k_B$.  We refer to Hoheisel 1996, for the estimated
physical values of the LJ parameters for several substances (as
an example, for a simple molecular fluid as N2, $\sigma \simeq 0.35$nm
and $\epsilon/k_B \simeq 100$K).

The rest of the computational domain (C) is described by the
Navier-Stokes equations. The fluid variables at C are the densities of
the conserved quantities for which the equations of motion in
conservative form are $\partial \Phi/\partial t=-\nabla \cdot
\mathbf{J}_{\Phi}$ with $\Phi=\left\{\rho, \rho\mathbf{u},\rho
e\right\}$ and $\mathbf{J}_{\Phi}=\left\{ \rho \mathbf{u},
\rho\mathbf{u}+\mathbf{\Pi}, \rho \mathbf{u} \,e + \mathbf{\Pi}\cdot
\mathbf{u} + \mathbf{q} \right\}$ standing for the mass, momentum and
energy fluxes respectively. Here $\rho$ is the density, $\mathbf{u}$
the local velocity, $e$ the specific energy,
$\mathbf{\Pi}=P\mathbf{1}+\mathbf{\tau}$ the stress tensor which
contains the pressure $P$ and the viscous tensor $\tau$ (for a
Newtonian fluid) and $\mathbf{q}=-\kappa\nabla\cdot T$, the heat flux
by conduction expressed via Fourier's law.  These continuum equations
may be solved via standard CFD methods.  Alternatively, for
low-Reynolds number flows ($Re \leq O(10)$) the equations can be
solved analytically (Delgado-Buscalioni \& Coveney 2003b), as is done
in the tests presented in \S \ref{s-cp}.

The kind of information to be transferred in the overlapping region
has been the subject of some discussion. The first attempts in the
literature (Delgado-Buscalioni \& Coveney 2003b and references therein)
considered the transfer of momentum in steady shear flows and proposed
a matching procedure based on the continuity of velocity across the
overlapping region. This sort of coupling strategy may be referred to as
``coupling-through-state''.  An alternative formulation of the
information exchange for liquids based on matching the fluxes of
conserved quantities (to/from P and C) was proposed by Flekk{\o}y {\em  et al.} 2000. 
These authors considered steady shear
flows with mass transfer.  In subsequent work by Delgado-Buscalioni \&
Coveney (2003b) the flux-coupling scheme was generalized to enable
transfer of mass, energy and momentum (along both transversal and
longitudinal directions).  Delgado-Buscalioni \& Coveney 2003b also
present a comparative study of the coupling-through-fluxes and coupling-through-state
schemes for flows involving energy transfer (longitudinal waves). It
was shown that the coupling-through-fluxes scheme provides the correct
physical behaviour, while the coupling-through-state scheme does not
guarantee positive entropy production. Consequently the coupling of fluxes is of central
importance in our hybrid scheme (see \S \ref{s-cp} and \S \ref{s-pc}).

\section{Temporal coupling}
\label{t-c}
In general there are three times involved in the coupling scheme: the
MD time-step $\Delta t_P$, the time-step for the C-solver $\Delta
t_C(>>\Delta t_P)$ and the averaging time $\Delta t_{av}$, which are
presented below as outline two possible strategies for
merging the time-evolution of C and P.  The information transfer
(from C$\rightarrow$P and P$\rightarrow$C) is updated over each
time interval, $\Delta t_C$.  As stated above, the P$\rightarrow$C
coupling consists firstly of a coarse-graining procedure. In
particular, for any particulate quantity, $\Phi_i$, the spatial average
over each P$\rightarrow$C cell of volume $V_{PC}$ ($=A \Delta X_{PC}$, in figure 1) is defined as
$\Phi(\mathbf{R},t)= \Sigma_{i\in V_{PC}} \Phi_i/N_{PC}$, where
$\mathbf{R}$ is the position of the cell in the coarse-grained
coordinates and $N_{PC}$ is the number of particles inside
$V_{PC}$. The time average also needs to be local with respect to the
coarse-grained dynamics. To that end, the microscopic quantities are
sampled over a time interval $\Delta t_{av}$ which is treated as an
independent parameter of the simulation:
\begin{equation}
\left<\Phi\right>({R},t_C)=
\frac{1}{\Delta t_C}\int_{t_C-\Delta t_av}^{t_C} \Phi(\mathbf{R},t)dt. 
\end{equation}
The magnitudes $\Delta t_C$ and $\Delta t_{av}$ are constrained by
several physical and numerical prerequisites quoted in Table 1.

\begin{table}
\begin{tabular}{l|lll}
Physical condition & \hspace{2.2cm} Algebraic & \hspace{-0.5cm}constraints   & Eq. \\
\hline
Local Equilibrium & $\Delta t_{C}>\tau_{col}= 0.14 \rho^{-1}T^{-1/2}$ & $\Delta x > \lambda= 0.2\rho^{-1}$ & C.1\\
Flow resolution & $\Delta t_{av}< O(0.1)\, \tau_{flow}$ & $|\Phi^{-1}\,d\Phi/dx|\Delta x<1$&       C.2 \\
Accuracy & $V_{PC} \Delta t_{av} > T/(\gamma^{2}\eta)$  & & C.3 \\
Courant condition & $\Delta t_C<\Delta x/(2 u_{flow})$ &  & C.4 \\
\end{tabular}
\caption{Constrains on the coarse grained time and length scales
within our hybrid MD-CFD scheme. Condition C.1 ensures
the local thermodynamical equilibrium at the averaging region: the
coarse-graining time $\Delta t_{av}$ and grid-spacing $\Delta x$ needs
to be larger than the collision time $\tau_{col}$ and the mean free
path $\lambda$, respectively. In C.1 $\tau_{col}$ and $\lambda$ are
estimated by the hard-sphere approximation. Condition C.2 is needed to
resolve the fastest flow characteristic time $\tau_{flow}$ and the
spatial variation of any physical variable $\Phi$ over the control
cell $|\Phi^{-1}\,d\Phi/dx|\Delta x$. Depending on the flow behaviour,
in C.2, $\tau_{flow}$ may stand for the period of the oscillatory flow
$f^{-1}$ or for the diffusive time $L_x^2/\nu$, etc.  The accuracy
condition in eq.(C.3) ensures that the signal-to-noise ratio of the
transversal momentum flux in a flow with shear rate $\gamma$ is
greater than one (similar kind of relationships can be derived for the
longitudinal momentum and energy fluxes).  The conditions C.1, C.2 and
C.3 are applied within the P$\rightarrow$C cell. The last condition,
C.4, ensures the stability of the numerical (explicit) scheme used for
time-integration of the C flow. The characteristic velocity of the
flow (on one grid space) is denoted by $u_{flow}$.}
\end{table}

There are essentially two ways to deal with the coupling of
time within the hybrid scheme: sequential coupling or synchronized coupling. The diagrams in
fig. 2 illustrate two possible choices for these time-coupling
strategies starting from given initial conditions. In the {\em
sequential coupling} scheme, both P and C are first moved to $t=\Delta t_C$
using the initial conditions. The C$\rightarrow$P coupling is
performed first at $t = \Delta t_C$ and the
P system is advanced to $t = 2 \Delta t_C \sim 300 \Delta t_P$.  
The averaged P-information collected over time interval $\Delta t_{av} = 2\Delta t_C$ 
within the P$\rightarrow$C cell is then
transferred to the C domain, giving the required boundary condition to
advance C towards the same time $t=2\Delta t_C$.  This procedure is
suited for serial processing.  More refined versions of sequential
coupling can be constructed to perform averaging over times $\Delta t_{av}$
greater than $\Delta t_{C}$.
 
In the {\em synchronized coupling} scheme both domains advance in time
independently until a certain instant at which both C$\rightarrow$P
and P$\rightarrow$C information transfers are exchanged.  This
scheme is suitable for parallel processing because
the P and C domains are being solved concurrently. We note that in this
case the averaged information from P transferred at any of these times
is obtained during the {\em previous} time interval $\Delta t_{av}$. This
fact introduces a delay of O($\Delta t_{av}/2$) in the C flow. 
Hence, it is important to ensure that $\Delta t_{av}$ is
about O$(10^{-1})$ times smaller than the fastest physical time of the flow
process (see Table 1).

\section{Continuum-to-particle coupling and its validation
\label{s-cp}}
The generalized forces arising from fluxes of mass, momentum and
energy measured from the C flow are to be injected into the particle system at the C$\rightarrow$P cell.
Table 2 summarizes how each flux contribution arising within C is
translated into the P domain.

Mass continuity is ensured by inserting or extracting particles at
a rate given by eq. (T.1) in Table 2. The convection of momentum is determined by
the product of the rate of particle insertion $s$ and the average velocity of the
incoming/outgoing particles $\langle \mathbf{v}^{\prime}\rangle$. By injecting
eq. (T.1) into eq. (T.2) it is  easily seen that convection balance requires
$\left<\mathbf{v^{\prime}}\right>=\mathbf{u}$.  
New particles are therefore introduced with velocities sampled from a
Maxwellian distribution at temperature $T$ and mean velocity
$\mathbf{u}$.  On the other hand, the local equilibrium 
$\left<\mathbf{\bar{v}}\right>=\mathbf{u}$ ensures that the average 
velocity of any extracted particles is equal to that of the continuum prescription

Viscous and pressure forces are introduced via external
forces acting on the particles at P$\rightarrow$C. 
An important issue is to decide how to distribute the overall force 
in eq. (T.3), $\sum^{N_{PC}} \mathbf{F}^{(ext)}_i$, over the
individual particles. We refer to Flekk{\o}y {\em et al.} (2000) and 
Delgado-Buscalioni \& Coveney (2003b) for a full discussion.
Although in general the force to be felt by each particle 
$i$ within the P$\rightarrow$C cell can be distributed according to
the particle positions (see Flekk{\o}y {\em et al.} 2000), we have adopted a flat
distribution $F^{ext}_i= A {\mathbf \Pi}\cdot {\mathbf n}/N_{PC}$
because it provides by construction, a correct rate of energy
dissipation in eq. (T.5) (see Delgado-Buscalioni \& Coveney 2003b).
Using eq. (T.1) it is seen that the balance of advected
energy in eq. (T.4) implies $\left<\epsilon^{\prime}\right>=e$.
The energy of each particle is composed of  kinetic and potential parts,
$\epsilon_i=v_i^2/2+\psi_i(\left\{\mathbf{r}\right\}^{N})$. The
specific energy of the continuum is $e=u^2/2+3kT/(2m)+\phi$ 
(here $\phi$ is the excess potential energy). The balance of kinetic energy
$\left<({v}^{\prime})^2\right>=u^2/2+3kT/(2m)$ is ensured by inserting
the new particles with the proper Maxwellian
distribution.  The balance of the potential energy requires a more difficult
condition $\left<\psi(\left\{\mathbf{r}\right\}^{N})\right>=\phi$ to be satisfied.
When inserting a new particle, this involves finding a precise location within the
C$\rightarrow$P cell with the desired potential energy.  To solve
this problem in a fast and effective way we have
constructed an algorithm for particle insertion called {\sc usher}
(Delgado-Buscalioni \& Coveney 2003c).
In order to find the site with the  desired energy within the complex 
potential energy landscape, the {\sc usher} 
algorithm uses a variation of the steepest descent algorithm including an
adaptable displacement. For densities within the range
$\rho=[0.4-0.8]$, the {\sc usher} scheme needs around $8-30$ 
iterations, each one involving the evaluation of a single-force. 
The {\sc usher} algorithm can be also applied in other problems involving particle insertion,
such as grand-canonical molecular dynamics.

Finally, eq. (T.6) in Table 2 determines
the rate of heat transfer into P by conduction. This energy can be
injected by reproducing a non-isothermal
environment within the C$\rightarrow$P cell. To that end we have implemented
a set of (typically 2-3) Nos\'e-Hoover
thermostats (NHT) separated by a distance $d$ with temperatures differing by $\Delta
T=[\nabla T\cdot \mathbf{n}]\,d$, where $\nabla T$ is the C-temperature gradient at C$\rightarrow$P.

\begin{table}
\begin{tabular}{p{2cm} p{2.5cm}|cccc}
Conserved quantity & Fluxes & P & $\leftarrow$ & C &  eq. \\
\hline
Mass &  & $m\mathrm{s}$ &=& $A \rho\mathbf{u} \cdot \mathbf{n}$ & (T.1) \\
 &  &  & &  &  \\
Momentum &
Convection & $m\mathrm{s}\left<\mathbf{v^{\prime}}\right>$ &=& $A\rho \mathbf{u u}\cdot {\mathbf n}$ & (T.2) \\
& Stress &
$\left<\sum^{N_{CP}} \mathbf{F}_i^{ext}\right>$ &=& $A {\mathbf \Pi}\cdot {\mathbf n}$ & (T.3) \\
 &  &  & &  &  \\
Energy &
Advection & $m\mathrm{s}\left<\epsilon^{\prime}\right>$ &=& $A\rho {\mathbf u} e\cdot \mathbf{n}$ & (T.4) \\
& Dissipation &
$\left<\sum^{N_{CP}} \mathbf{F}_i^{ext}\cdot \mathbf {v}_i\right>$ &=& $A\mathbf{\Pi\cdot u}\cdot \mathbf{n}$ & (T.5) \\
& Conduction & $\left<\mathbf{J}_Q^{ext}\right>\cdot \mathbf{n}$ &=& $A\mathbf{q}\cdot \mathbf{n}$ & (T.6) \\
\hline
\end{tabular}
\caption{The balance of mass, momentum  and energy fluxes at each
  $C\rightarrow P$ cell. The fluxes measured within C (third column) are
imposed into P via the expressions given at the second column.
The cell's surface is $A$, and the surface vector $\mathbf{n}$ points
outwards (fig. 1).  The mass rate 
is $\mathrm{s}(t)$ (s$>$0 for inserted and s$<$0 for removed particles).
The velocity  and energy  of  the inserted/removed  particles
are $\mathbf{v^{\prime}}$ and $\epsilon^{\prime}$ respectively.
The external force and heat flux inserted within the C$\rightarrow$P
cell are $\sum^{N_{PC}} F^{ext}_i$  and
  $\left<\mathbf{J}_Q^{ext}\right>$, respectively.}
\end{table}

The decay of transversal and longitudinal waves is an excellent test
for the validity of our proposed the C$\rightarrow$P coupling as they comprise the whole set of
hydrodynamic modes: shear, sound and heat waves.  For these tests we
implemented a set-up consisting of a P region of length $L_x$ (with
periodic boundary conditions in $y$ and $z$ directions) surrounded by
two C domains.  We initially imposed on the P system a sinusoidal (x- or
y-) velocity profile along the x direction. By extracting the initial
amplitudes of the spatial Fourier components of all the hydrodynamic
quantities it is then possible to trace the entire time-evolution of the 
relaxing flow using linear hydrodynamics. In particular, this permits us
to calculate at any time the generalized forces to be inserted within
the C$\rightarrow$P cell.  The time evolution of the spatial Fourier
components of the P-variables is finally compared with the analytical
expressions.  Such kinds of comparisons are shown in figures 3 and 4,
for the case of a relaxing shear wave and a longitudinal
wave, respectively.  The excellent agreement obtained indicates that the C$\rightarrow$P
coupling protocol can be used for capturing fast and low-amplitude
flows, such as those governed by sound, shear or heat waves.

The entropy perturbation, shown in fig. 5, was calculated from the
temperature and density perturbative field. The results clearly show
that using only one thermostat per C$\rightarrow$P cell (denoted by
1-NHT$_{CP}$, in figs. 3 and 4) leads to negative entropy production.
The pure exponential decay of heat due to diffusion is only
recovered when the correct (averaged) heat flux is connected to
each C$\rightarrow$P cell;  in figs. 4 and 5 we present a result with two thermostats per cell (2-NHT$_{CP}$).  
This result confirms that the coupling-through-fluxes scheme is
the correct matching procedure.

\section{Particle-to-continuum coupling: finite volumes and fluctuations
\label{s-pc}}

Within the P$\rightarrow$C cells
the information coming from the
particle dynamics is coarse-grained to provide boundary conditions at the ``upper''
C-level. In \S \ref{s-o} we introduced the 
averages needed to produce such information. At the P$\rightarrow$C interface
the C region receives the averaged particle-fluxes as
open-flux (von Neumann) boundary conditions. The averaged mass, momentum and energy 
particle-fluxes through the P$\rightarrow$C interface are constructed
as follows,
\begin{eqnarray}
\rho\mathbf{u}\cdot\mathbf{n}_{PC}&=&\frac1V_{PC}\left<\Sigma_{i=1}^{{N}_{PC}}
m \mathbf{v}_i\right>\cdot\mathbf{n}_{PC} \\
\mathbf{\Pi}\cdot\mathbf{n}_{PC}&=&\frac1V_{PC} \left< \left(\Sigma_{i=1}^{{N}_{PC}} m \mathbf{v}_i\mathbf{v}_i
-\frac12\Sigma_{i,j}^{{N}_{PC}} \mathbf{r}_{ij}\mathbf{F}_{ij}\right)\right>\cdot\mathbf{n}_{PC}\\
\mathbf{q}\cdot\mathbf{n}_{PC}&=&\frac1V_{PC} \left<\left(\Sigma_{i=1}^{{N}_{PC}} \epsilon_i\mathbf{v}_i
-\frac12\Sigma_{i,j}^{{N}_{PC}} \mathbf{r}_{ij} \mathbf{v}_i\mathbf{F}_{ij}\right)\right>\cdot\mathbf{n}_{PC}
\end{eqnarray}
where $N_{PC}$ is the number of particles inside the P$\rightarrow$C
cell of volume $V_{PC}$ and $\mathbf{n}_{PC}$ 
is the surface vector shown in fig. 1.

\subsection{Hybrid finite volume: boundary conditions}

Let us now illustrate how these fluxes can be injected into the C domain in the
framework of the finite volumes method (Patankar 1980). The finite volumes
method is ideally suited to our scheme because it exactly balances the fluxes
across the computational cells. Its principle is simple.
Briefly, the computational domain (C) is divided into cells
of volume $V_{l}$ whose size and location is given by the nodes of a
specified mesh, $\left\{ \mathbf{R}_{l}\right\} $, $l=\left\{
1,...,M_{c}\right\} $.  Integrating the conservation equation
$\partial \Phi/\partial t=-\nabla \cdot \mathbf{J}_{\Phi}$  over each
computational cell (say the cell $H$ in fig.  1) one
obtains,\begin{equation} \frac{d\, \rho _{H}\Phi
_{H}}{dt}=\frac{1}{V_{H}}\sum _{f}A_{f}\mathbf{J}_{\Phi ,f}\cdot
\mathbf{n}_{f}.
\label{int.vol}
\end{equation}
where $A_{f}$ stands for the area of the face $f$ and ${\bf n}_{f}$ is
the outwards normal surface vector. The volume integral of the
transient term of the conservation equation has been approximated by
$V_{H}$ times the explicit time derivative of the value of the
integrand at the cell centre, halfway between the surfaces: $\rho_H
\Phi_H$.  Equation (\ref{int.vol}) yields a set of ordinary
differential equations (ODE's) involving the flow variables at each
cell face, $f$. The set of equations is closed for the flow variables
at the cell centre by expressing the fluxes at the interfaces
$\mathbf{J}_{\Phi,f}$ in terms of differences of flow variables at
neighbouring cell centres, via the constitutive relations.

Let us consider the momentum flux balance for the low Reynolds number
flow of an incompressible and isothermal fluid driven by diffusion of
$y$-velocity along $x$ direction: ${\bf u}=u(x){\bf j}$.  In this
case ${\bf J}\cdot{\bf n} =P\,{\bf i}- \eta (du/dx) {\bf j}$,
where the surface vector of the P$\rightarrow$C surface is ${\bf n}={\bf i}$.
Let us consider an isobaric environment 
and restrict ourselves to the transfer of transversal ($y$)
momentum, governed by the momentum flux $J \equiv {\bf J}\cdot{\bf
j}=-\eta \gamma $ and the shear rate $\gamma \equiv du/dx$. 
Integrating along the cell $H$ (see fig. 1), using a first
order space discretisation of the stress
(e.g. $J_w=-\eta(u_H-u_W)/\Delta x$) and an explicit time integration
scheme, one obtains
\begin{equation}
\label{exp}
u_{H}^{n} =u_{H}(1-2r)+ru_{E}+ru_{W},
\end{equation}
where the subscripts $H$ denote the set of cell centres
$H=\left\{1,M\right\}$, and the symbols $E$ (east) and $W$ (west)
denote variables measured: $x=E(=H+1)$ and $x=W(=H-1)$. The time
instant is denoted by $u_{H}=u(x_{H},t)$ and
$u_{H}^{n}=u(x_{H},t+\Delta t_{C})$ and $r\equiv \nu \Delta t/(\Delta
x^2)$.  with $\nu=\eta/\rho$ the kinematic viscosity.  In order to
guarantee the numerical stability of the explicit scheme in
eq. (\ref{exp}), the size of the (smallest) control cell inside the C
region $\Delta x$ and the time step $\Delta t$ are related through
$r\leq 1/2$, which corresponds to the grid-diffusive-velocity
$u_{flow}=\nu/\Delta x$ in the Courant condition C.4 of Table 1.  In
solving eq. (\ref{exp}) we used a uniform grid with a typical value of
$\Delta x \sim 0.5$.

In order to impose the boundary condition one needs to determine the
velocity within the outer cells: at the rightmost 
$x=x_{M+1}=L_x$ and at the leftmost boundary (inside the P$\rightarrow$C
cell, see fig. 1) $x=x_0=l_C-\Delta x/2$. At $x_{M+1}=L_x$ there is a rigid wall which moves at a velocity
$u_{wall}(t)$ and provides the Dirichlet boundary condition $u_{M+1}=u_{wall}(t)$.
The hybrid formulation is applied at the left boundary
$x_{0}=l_{C}-\Delta x/2$. To evaluate the outer velocity $u_{W}=u_0$ we impose the
balance of momentum flux across the $w$ surface at $x=l_C$. This means
that  the continuum flux evaluated at $x=w$ is made equal to the corresponding
averaged particle flux $\langle j\rangle_{w}=-\eta ({\bar{u}_{H}}-u_{W})/\Delta x$. 
The outer velocity to be inserted in eq. (\ref{exp}) is then
$u_{W}={\bar{u}_{H}}+\langle j\rangle_{w}\Delta x/\eta $. The velocity $\bar{u}_{H}$ 
is evaluated as a linear combination of the continuum
$u_{H}(=u_1)$ and the average particle velocity $\langle
v\rangle_{H}$ at $x_1=l_C+\Delta x/2$:
\begin{equation}
\label{upb}
{\bar{u}_{H}}=(1-\alpha )u_{H}+\alpha \langle v\rangle_{H}.
\end{equation}
By inserting eq. (\ref{upb}) into eq. (\ref{exp}) one obtains
the velocity at the boundary cell 
\begin{equation}
\label{up}
 u_{H}^{n}=u_{H}\,(1-r)+r\,u_{E}
+\frac{\langle j\rangle_{w}\Delta t}{\rho \Delta x_{H}}
+\alpha r\,\left(\langle v\rangle_{H}-u_{H}\right).
\end{equation}
The reason for the choice of $\bar{u}_{H}$ in eq. (\ref{upb}) now becomes
clear. It introduces the last term on the right hand side of eq. (\ref{up})
which acts as a forcing term ensuring velocity continuity by gently driving the continuum velocity
to the corresponding particle average $u_{H}=\langle v\rangle _{H}$.
The strength of the {\em velocity coupling} is maximal when $\alpha =1$ and is
absent if $\alpha =0$. The idea of using a hybrid gradient (arising
for any $\alpha\ne 0$ in eq. (\ref{upb})) arose from the outcome of
calculations performed at very low shear rates ($\gamma<10^{-2}$).  Using
$\alpha=0$ one obtains a velocity discontinuity at P$\rightarrow$C
which is of the same order of magnitude as the fluctuations of the mean
instantaneous velocity within the overlapping region.  At low shear rates
this means substantial relative differences in the C and P velocities,
$<v>_{H}-u_{H})/u_{H}\sim O(1)$.  This problem is solved by introducing
a small velocity coupling in the continuum scheme, with a small value
of $\alpha \in [0.2,0.5]$, which drives the continuum velocity to the
average particle velocity in a time of $O[\Delta x^{2}/(\nu \alpha)]$.
To check any influence of the velocity coupling term in eq. (\ref{up}) 
on the flux balance, we performed
simulations of the Couette flow at different shear rates and compared its average over time
with the time averaged momentum particle flux. 
The results showed that, in average, the velocity coupling term is 
vanishingly small so it does not introduce any extra flux in the coarse-grained time scale.

\subsection{The effect of fluctuations: shear stress}
In our scheme, the fluctuating nature of the fluxes introduced into
the C region at P$\rightarrow $C imposes a limitation on our ability
to resolve the flow field, as 
also arises in experiments and full MD simulations. This limit is
determined by signal-to-noise ratio becoming smaller than one. 
A theoretical expression for the amplitude of the stress
fluctuations can be obtained (Delgado-Buscalioni {\em et al.} 2003),
providing a relationship between the signal-to-noise ratio 
and the coarse-grained time and space scales $\Delta t_{av}$ and
$V_{PC}$. Table 1 contains the condition
to ensure an averaged shear force larger than its variance.
It is clear that in weak steady flows it is always possible to increase the
signal-to-noise ratio by enlarging $\Delta t_{av}$.  Nevertheless,
in a general space and time-dependent flow, the sizes of the averaging windows in space and time
($V_{PC}$ and $\Delta t_{av}$) are bounded above by the minimum
wavelength and characteristic time which need to be treated
within the flow. Such requirements on spatial and temporal flow resolution are also quoted in Table 1.

\section{Oscillatory wall flow
\label{osc}}
In order to test the applicability of the full hybrid scheme under
unsteady flows, we have considered the flow of an incompressible and
isothermal fluid between two parallel walls in relative oscillatory
motion. This set-up is widely used to investigate the reological
properties of complex fluids attached to surfaces, as polymer brushes
(see C.M. Wijmans \& B. Smit (2002) for a recent review). These
systems are good examples of the sort of applications of the hybrid
scheme, which can treat the complex fluid region by MD and the
momentum transfer from the bulk by CFD. A similar set-up can be also
used in the simualation of nanotechnological process. For instance,
Stroock {\em et al.} (2002) showed that the mixing of
solutions in low Reynolds number flows in microchannels
can be enhanced by introducing bas-relief nano-structures on the
floor of the slot. In our test flow, the simulation domain is
$0\leq x\leq L_x$ and it is periodic along $y$ and $z$ directions.
The particle domain occupies the region $x<l_P$, and it includes 
the LJ liquid and the atomistic wall composed of two
layers LJ particles at $x\leq 0$. The
continuum domain comprises the region $x\in[l_C,L_x]$. The
sizes of the simulation domains were within the nanoscale  
$L_x \sim 50\sigma$, and $l_P \sim 15\sigma$, 
while the width of the overlapping region, $l_P-l_C$, was
set to arround $5\sigma$.  The flow is uniquely driven by the oscillatory
motion of the $x=L_x$ wall along the $y$ direction, meaning that the mean pressure
is constant throughout the domain and there are no transfers of mean
energy or mass in the $x$ direction (perpendicular to the
P$\rightarrow$C surface). Therefore the mean flow carries
transversal momentum by diffusion only, and the equation of motion for the
$y$-velocity is $\partial u/\partial t=\nu \partial^2 u/\partial x^2$,
with boundary conditions $u(0,t)=0$ and
$u(L,t)=u_{wall}(t)=u_{\max}\,\sin(\omega t)$.  This equation can be
solved analytically (H. Schliting 1958; C.M. Wijmans \& B. Smit 2002).
The flow profile has a maximum amplitude at the moving wall and the
momentum introduced by its motion penetrates into a fluid layer of
width $\delta \sim \sqrt{\pi\nu/f}$.  Beyond this layer the flow
amplitude tends to zero diffusively as it approaches the other wall held
at rest. Therefore, the maximum shear rate attained inside the
momentum layer is of order $\gamma \sim u_{\max}/\delta$.  Inserting
this relation into the signal-to-noise condition (C.3 in Table 1), we
find
\begin{equation}
\rho u_{\max}^2 \Delta t_{av} > \pi f^{-1} \left(\frac{k_B T}{V_{PC}}\right).
\label{s3}
\end{equation}
Equation (\ref{s3}) means that in order to attain a signal-to-noise
ratio larger than one, the mean kinetic energy per unit volume of the
flow integrated over the averaging time $\Delta t_{av}$ needs to be
larger than the corresponding energy due to fluctuations over the
period of the mean flow. It is important to mention that at low enough
frequencies ($f>\nu/L^2$), there is sufficient time for momentum to be
spread by diffusion over the whole domain. In such situations the
correct condition is given by the signal-to-noise condition (C.3 in
Table 1) with $\gamma\sim u_{\max}/L^2$.

As indicated by condition C.2 in Table 1, in order to solve for the
temporal variation of the flow it is required that $\Delta t_{av} f
\geq O(0.1)$. Inserting this condition into eq. (\ref{s3}) one obtains
$u_{\max} > 5 \left(\frac{k_B T}{\rho V_{PC}}\right)^{1/2}$.  For $k_B
T=1.0$, $\rho=0.8$ and $V_{PC}=O(100)$ the above inequality yields
$u_{\max}> 0.5$. We performed oscillatory shear simulations for values
of $u_{max}$ above, close to and below the threshold given by
eq. (\ref{s3}). As shown in fig. 6a, calculations made at large flow
amplitudes are in excellent agreement with the analytical solution.
In figure 6b we present results for the same density and temperature
($\rho=0.8$ and $T=1$) and a wall velocity $u_{\max}=0.5$ right at the
accuracy limit predicted by (\ref{s3}).  The averaging time was chosen
to be $\Delta t_{av} =10$.  As shown by the instantaneous velocity within
the P$\rightarrow$C cell, the noise amplitude is nearly equal to the
flow amplitude and its time-averaged value shows traces of
fluctuations. Figure 6c corresponds to the same velocity and density
but at a larger temperature $T=4$. This case is below the accuracy
limit (given by C.3 in Table 1) where forces arising from thermal
fluctuations dominate the hydrodynamic ones.

\section{Conclusions and future directions
\label{cn}
}

We have presented a hybrid continuum-particle scheme for
moderate-to-large fluid densities which takes into account mass,
momentum and energy exchange between a domain described by discrete
particle Newtonian molecular dynamics (P) and an interfacing domain
described by continuum fluid dynamics (C). The coupling scheme is
applied within an overlapping region comprised of two sub-cells where
the two-way exchange of information is performed: C$\rightarrow$P and
P$\rightarrow$C. We have shown that the coupling-through-variables
scheme (which simply ensures continuity of variables within the
overlapping region) is not sufficient to guarantee positive entropy
production.  However, by generalizing the coupling-through-fluxes
scheme proposed by Flekk{\o}y {\em et al.}, 2000 to energy and mass
transfer we find that the correct decay of shear, sound and heat waves
is obtained.

We are now deploying the present scheme to study the dynamics of a
tethered polymer under shear flow. The polymer and its local
environment are treated via MD, while the shear flow imposed on the
outer domain is treated via the finite volume CFD method.  In the
future, we plan to apply our hybrid scheme to the study of membrane
dynamics.

Enhancements to the present hybrid algorithm are under
investigation. In the scheme described here the energy flux balance is
ensured only over time averages.  We are currently studying
alternative schemes which {\em exactly} balance this flux. From a
numerical standpoint, we plan to implement the P$\rightarrow$C
coupling in conjunction with a finite volume CFD solver in 3D. 

Also, the present scheme can be easily adapted to couple  molecular dynamics
with another mesoscopic scheme that takes into account hydrodynamic
fluctuations. This sort of hybrid scheme could be used in applications
where the fluctuations are relevant (microfluidics, fluids near
critical point, etc...). An important condition for the interfacing mesoscopic scheme is that
it needs to be fully consistent with thermodynamics. Also
important is that the transport coefficients of the mesoscopic model should be adjustable to
represent the correct coarse-grained dynamics of the selected working
fluid. Natural candidates are the Lagrangian schemes 
involving Voronoi tesselation (Flekk{\o}y {\em et al.} 2000a) or
the Smooth Particle Dynamics model and related mesoscopic techniques (Espa\~nol 2003).
The lattice Boltzmann (LB) method is another possible candidate to
interface with the MD domain. This model has been already used in multiscale modelling
(Succi {\em et al} 2001).
Nevertheless, the problem with LB methods at present is that there is no truly reliable 
thermohydrodynamic model other than for single phase flow. 
Energy  conservation remains unsolved and most models are athermal; even the
thermohydrodynamic lattice-BGK models for the ideal gas are vastly 
over-determined and get the temperature dependence of the viscosity wrong
(Boghosian and Coveney 1998). 
Therefore the hybrid scheme proposed here could only be interfaced with 
the lattice Boltzmann model in certain applications involving
isothermal and incompressible single phase flows.

A longer term goal of this research is to develop a flexible,
componentized, hybrid coupling environment into which {\em any}
molecular dynamics and {\em any} continuum fluid dynamics codes may be
inserted. This will require consideration of electrostatic forces and,
therefore, an additional conserved quantity, the electric charge,
whose flux coupling will requires use of Poisson-Boltzmann
solvers. Moreover, such multiscale hybrid schemes are attractive
candidates for efficient deployment on computational grids, a feature
now under investigation with the RealityGrid project
(www.realitygrid.org).

\section{Acknowledgements}
We gratefully acknowledge fruitful discussions with Professor Eirik
Flekk{\o}y. This research is supported by the European Commission
through a Marie Curie Fellowship to RD-B (HPMF-CT-2001-01210) and by
the EPSRC RealityGrid project GR/R67699. R. D-B also acknowledges
support from the project BFM2001-0290.

\section*{References}
\begin{itemize}
\item Abraham F. F., Broughton J. Q., Berstein N. and Kaxiras E. 1998
  Spanning the continuum to quantum length scales in a dynamic
  simulation of brittle fracture. {\em Europhys. Lett.}, {\bf 44} 783
\item Boghosian B. and Coveney P. V 1998 
Inverse Chapman-Enskog derivation of the thermohydrodynamic lattice-BGK model for the ideal gas, {\em Int. J. Mod. Phys. C} {\bf 9}, 1231-1245
\item Delgado-Buscalioni \& Coveney P. V. 2003a Continuum-particle hybrid
	coupling for mass, momentum and energy transfers in unsteady
	fluid flow, {\em Virtual Journal of NanoScale Science \& Technology}
         {\bf 7}, Issue 16, April 21, http://ojps.aip.org/nano
\item Delgado-Buscalioni R. \& Coveney P. V. 2003b Continuum-particle
  hybrid coupling for mass, momentum and energy transfers in unsteady
  fluid flow, {\em Phys. Rev. E} {\bf 67}, 046704.
\item Delgado-Buscalioni R. \& Coveney P. V. 2003c {\sc Usher}: an
  algorithm for particle insertion in dense fluids.  {\em  J. Chem. Phys} {\bf 119}, 978.
\item Delgado-Buscalioni R., Coveney P. V. and Flekk{\o}y E. 2003,   
Oscillatory shear flow in liquids via hybrid continuum-particle
scheme, submitted to {\em Phys. Rev. E} 
\item  Espa\~nol P. 2003  Dissipative Particle Dynamics, in Trends
     in {\em Nanoscale Mechanics: Analysis of Nanostructured Materials
     and Multi-Scale Modeling}, V. M. Harik and M. D. Salas editors
     (Kluwer 2003).
\item Flekk{\o}y E., Wagner G. and Feder J. 2000 Hybrid Model for
Combined Particle and Continuum Dynamics.  {\em Europhys. Lett.} {\bf
52}(3) 271-276.
\item Flekk{\o}y E, P.V. Covney and G. De Fabritiis 2000a, 
Foundations of dissipative particle dynamics, {\em Phys. Rev. E} {\bf
  62}, 2140
\item Garcia A., Bell J., Crutchfield Y. and Alder B. 1999 Adaptive
  Mesh and Algorithm Refinement using Direct Simulation Monte
  Carlo. {\em J. Comp. Phys.}, {\bf 154}, 134.
\item Hoheisel C. 1996, Computer Calculation, in {\em Transport
  properties of fluids: their correlation, prediction and estimation},
  H. Millat, J.J. Dymomd and C.A. Nieto de Castro eds., Cambridge
  University Press.
\item  Schliting H. 1958 {\em Grenzchicht-Theory}, Braun ed.,
  Karlsruhe.
\item Stroock A. D., S. K. W.  Dertinger, A. Ajdar, I. Mezi\'c,
  H. A. Stone, G. M. Whithesides 2002  Chaotic mixer for microchannels,
  {\em Science} {\bf 295}, 647.
\item Patankar S. 1980, {\em Numerical Heat Transfer and Fluid Flow},
Hemisphere, New York.
\item Succi et al 2001 Applying the lattice Boltzmann equation to
                 multiscale fluid problems, {\em Computers in Sci. and Eng.}  {\bf 3}, 26-37
\item  Wijmans C.M. \&  Smit B. 2002 Simulating thethered polymer layers
in shear flow with dissipative particle dynamics. {\em Macromolecules}
{\bf 35}, 7138-7148.
\end{itemize}

\newpage

\begin{figure}
\includegraphics[width=13cm,totalheight=5cm]{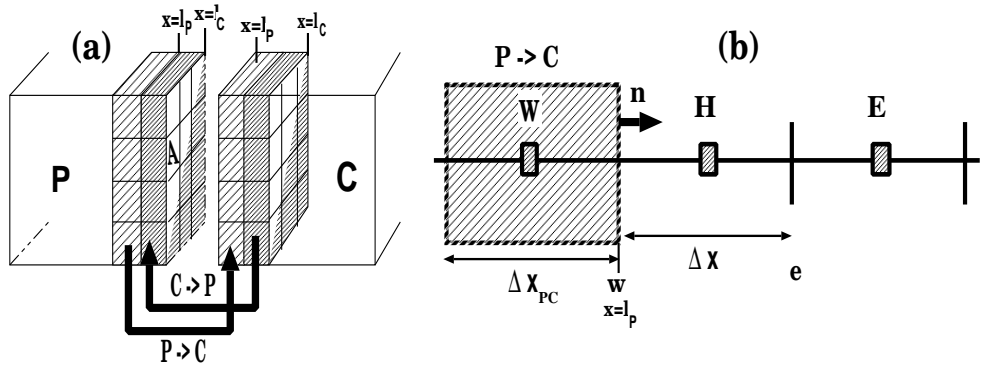}
\caption{The domain decomposition of the hybrid scheme: (a) displays P
and C regions separatedly. The shaded region represents the
overlapping domain comprised by a 2D array of C$\rightarrow$P and
P$\rightarrow$C cells where the exchange of microscopic and
macroscopic information is carried out. The surface area of each cell
is $A$. (b) shows the P$\rightarrow$C region in more detail and the
neighbouring control cells pertaining to the finite volume
discretization of the C region. In this one-dimensional example, the
width of the P$\rightarrow$C cell is $\Delta x_{PC}$ and its volume is
$V_{PC}=A\Delta x_{PC}$.}
\end{figure}

\begin{figure}
\includegraphics[width=8cm,totalheight=5cm]{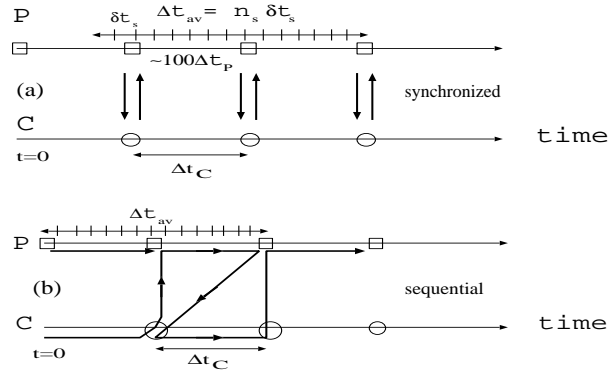}
\caption{ Two possible time coupling strategies in a
particle-continuum hybrid scheme: (a) synchronized coupling and (b)
sequential coupling.  Bold arrows indicate the direction of the
information transfer.  The time average of the P variables is
performed during the time interval $\Delta t_{av}$ by $n_s$ samplings
separated in time by $\delta t_s$.  $\Delta t_C$ and $\Delta t_P$ are
the continuum time step and the MD time step respectively.}
\end{figure}

\begin{figure}
\includegraphics[width=12cm,totalheight=4cm]{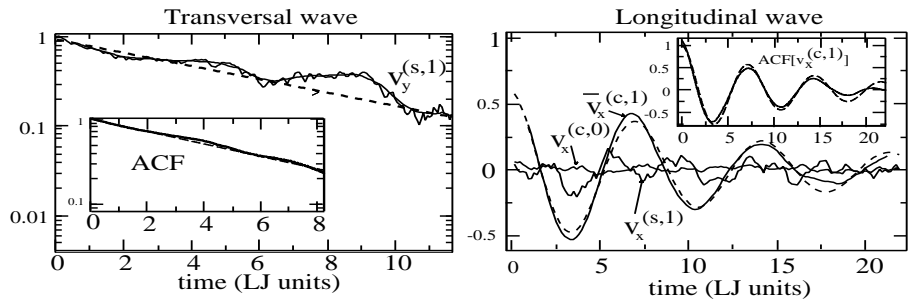}
\caption{ The Fourier components of the transversal and longitudinal
velocity perturbation.  In the notation $v_y^{(s,n)}$; $s$ indicates
the sinusoidal component and $c$ cosinusoidal and $n$ the wavenumber
$k_n=nk_0$.  The transversal wave has $k_0=0.35$, the size of the P
region was and $L_x=20$ (in LJ-units) and the temperature was $T=2.5$;
while for the longitudinal wave $k_0=0.168$, $L_x=40$ and $T=3.5$. In
both cases $\rho=0.53$ The autocorrelation of the velocity is also
shown. In all graphs the dashed lines are the analytic solution from
linear hydrodynamics. Reproduced from R. Delgado-Buscalioni \& Coveney
2003b with permision}
\end{figure}

\begin{figure}
\includegraphics[width=12cm,totalheight=6cm]{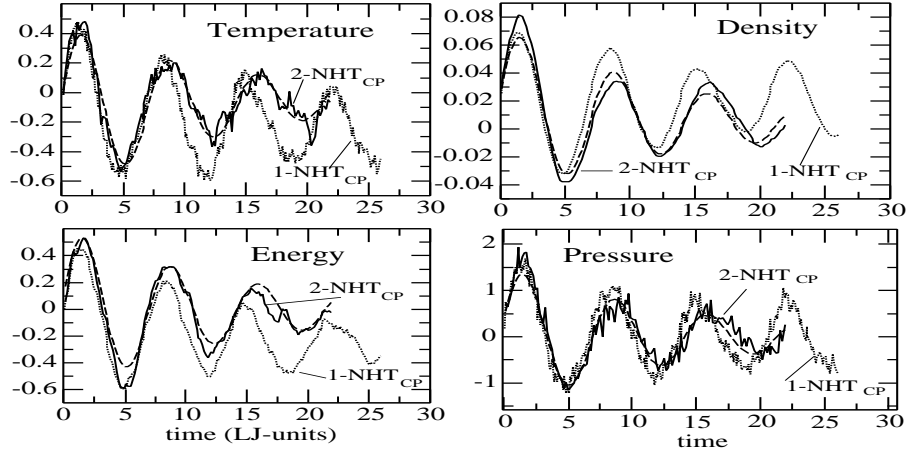}
\caption{ The dominant Fourier mode of the various thermodynamic
variables in the decay of the same longitudinal wave shown in figure
3.  Comparison is made between a calculation with tow Nos\'e-Hoover
thermostats per C$\rightarrow$ P cell (2-NHT$_{CP}$) and another using
only one thermostat (1-NHT$_{CP}$).  Dashed lines are the analytical
hydrodynamic solution.  The entropy production from these two
simulations is shown in figure 5, only the one with two thermostats
yields th correct physical behaviour. Reproduced from
R. Delgado-Buscalioni \& Coveney 2003b with permision}
\end{figure}

\begin{figure}
\includegraphics[width=6cm,totalheight=4cm]{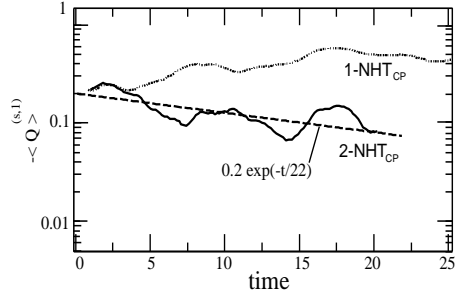}
\caption{The main Fourier mode of the entropy density (as a product
  with the mean temperature) $-\left<Q^{(s,1)}\right>$ time-averaged
  along $\Delta t_{av} =1.0$. The result comes from the same
  longitudinal wave shown in figs. 3 and 4.  Comparison is made
  between a flux-coupling scheme (using 2-NHT$_{CP}$) and the
  coupling-state scheme using 1-NHT$_{CP}$ (cf. fig. 4). The latter
  violates the second law of thermodynamics. The dashed line is the
  analytical hydrodynamic result.  Reproduced from
  R. Delgado-Buscalioni \& Coveney 2003b with permision.}
\end{figure}

\begin{figure}[h]
\includegraphics[width=12cm,totalheight=7cm]{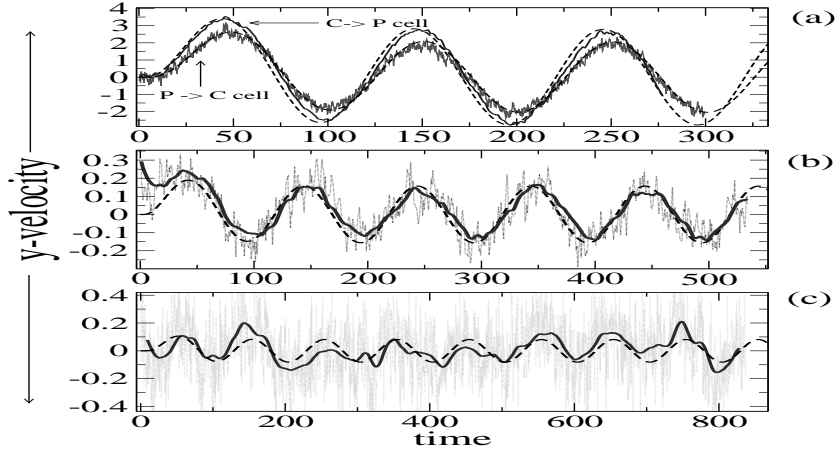}
\caption{Mean molecular velocities within the overlapping region for
several oscillatory-wall shear flows applied to a LJ fluid; $u_{\max}$ is the
maximum wall  velocity and $f$ its frequency.  (a) Flow corresponds to
$u_{\max}=10$, $f=0.01$ and $T=1.0$; we plot the
instantaneous particle velocity at P$\rightarrow$C and the
time-averaged particle velocity (along $\Delta t_{av} =1$) at
C$\rightarrow$P; (b)
corresponds to $u_{\max}=0.5$, $f=0.01$ and $T=1.0$; (c) to
$u_{\max}=0.5$, $f=0.01$ and $T=4.0$. In all cases $\rho=0.8$, the
extent of the periodic directions are $L_y=L_z=9$, while
$V_{PC}=\Delta x\,L_y\,L_z=178$.  In (b) and (c) we show the
P$\rightarrow$C mean velocity (instantaneous and time-averaged velocity with
$\Delta t_{av}=10$); dashed lines are the analytical hydrodynamic solutions of the
imposed shear flows. All quantities are given in reduced LJ units.}
\end{figure}

\end{document}